\documentclass[reprint,pre,twocolumn,showpacs,superscriptaddress]{revtex4-2}
\usepackage{amsmath,amsthm,amssymb,amscd,float,graphicx,enumerate,float,bbold}
\usepackage[utf8]{inputenc}
\usepackage[T1]{fontenc}
\usepackage{lmodern}
\usepackage[colorlinks,linkcolor=blue,citecolor=blue,urlcolor=blue]{hyperref}
\newcommand{\al}{\alpha}
\newcommand{\be}{\beta}
\newcommand{\de}{\delta}

\newcommand{\vep}{\varepsilon}
\newcommand{\ga}{\gamma}

\newcommand{\la}{\lambda}

\newcommand{\si}{\sigma}

\newcommand{\vp}{\varphi}

\newcommand{\ze}{\zeta}
\newcommand{\De}{\Delta}

\newcommand{\bw}{\mathbf{w}}

\newcommand{\bv}{\mathbf{v}}

\newcommand{\bsi}{{\boldsymbol{\si}}}

\newcommand{\tS}{\widetilde{S}}

\newcommand{\CC}{{\mathbb C}}

\newcommand{\cE}{{\mathcal E}}

\newcommand{\cP}{{\mathcal P}}

\newcommand{\pd}{\partial}

\newcommand{\ket}[1]{|#1\rangle}

\newcommand{\mss}{\kern 1pt}

\renewcommand{\le}{\leqslant}
\renewcommand{\ge}{\geqslant}
\newcommand{\tends}[1]{\bbuildrel{\hbox to 2em{\rightarrowfill}}_{#1}^{}}

\newcommand{\csch}{\operatorname{csch}}

\newcommand{\Li}{\operatorname{Li}}
\newcommand{\tr}{\operatorname{tr}}

\newcommand{\iu}{\mathrm i}
\newcommand{\diff}{\mathrm{d}}

\newcommand{\su}{\mathrm{su}}

\newcommand{\sla}{\mathrm{sl}}

\newcommand{\en}{\enspace}

\newcommand{\Int}[1]{\,\mathop{\!#1}\limits^{\lower1ex\hbox{$\scriptstyle\circ$}}{}}

\theoremstyle{remark}
\newtheorem{remark}{Remark}

\newcommand{\bsv}{\mathbf s}

\newcommand{\cV}{c_{\mathrm{V}}}
\def\clap#1{\hbox to 0pt{\hss#1\hss}}

\begin{document}

\title{Thermodynamics and criticality of $\su(m)$ spin chains of Haldane--Shastry type}

\author{Federico Finkel}\email{ffinkel@ucm.es}
\author{Artemio González-López}\email[Corresponding author. Email address: ]{artemio@ucm.es}

\affiliation{Depto.~de Física Teórica, Facultad de Ciencias Físicas,\\
  Universidad Complutense de Madrid,\\
  Plaza de las Ciencias 1, 28040 Madrid, SPAIN}
\date{\today}
\begin{abstract}
  We study the thermodynamics and critical behavior of $\su(m)$ spin chains of Haldane--Shastry
  type at zero chemical potential, both in the $A_{N-1}$ and $BC_N$ cases. We evaluate in closed
  form the free energy per spin for arbitrary values of $m$, from which we derive explicit
  formulas for the energy, entropy and specific heat per spin. 
  In particular, we find that the specific heat features a single Schottky peak, whose temperature
  is well approximated for $m\lesssim10$ by the corresponding temperature for an $m$-level system
  with uniformly spaced levels. We show that at low temperatures the free energy per spin of the
  models under study behaves as that of a one-dimensional conformal field theory with central
  charge $c=m-1$ (with the only exception of the Frahm--Inozemtsev chain with zero value of its
  parameter). However, from a detailed study of the ground state degeneracy and the low-energy
  excitations, we conclude that these models are only critical in the antiferromagnetic case, with
  a few exceptions that we fully specify.
\end{abstract}

\maketitle


\section{Introduction}\label{sec.intro}

In this paper we shall consider a broad class of spin chains with long-range interactions modeled
on the Haldane--Shastry chain~\cite{Ha88,Sh88}, whose interactions can be expressed in terms of
the generators of the $\su(m)$ algebra in the fundamental representation. More precisely, if $N$
denotes the number of sites and $m$ is the number of internal degrees of freedom the canonical
basis of the system's Hilbert space $\otimes_{i=1}^N\CC^{m}$ is spanned by the vectors
\begin{equation}\label{basis}
  |s_1,\dots, s_N\rangle:=|s_1\rangle\otimes\cdots\otimes|s_N\rangle,
\end{equation}
where $s_i\in\{1,\dots,m\}$. We define the permutation and spin flip operators $S_{ij}$ and $S_i$
($1\le i<j\le N$) by the usual formulas
\begin{align}
  \label{Pij}
  S_{ij}\ket{\cdots s_i\cdots s_j\cdots }&=\ket{\cdots s_j\cdots s_i\cdots},\\
  S_{i}\ket{\cdots s_i\cdots }&=\ket{\cdots m-s_i+1 \cdots}.
\end{align}
The latter operators can be easily expressed in terms of the (Hermitian) $\su(m)$ generators
$T^\al_k$ in the fundamental representation, where $1\le \al\le m^2-1$ and the subindex labels the
chain sites. Indeed, using the normalization $\tr(T_k^\al T_k^\be)=2\de_{\al\be}$ we have
\[
  S_{ij}=\frac1m+\frac12\sum_{\al=1}^{m^2-1}T_i^\al T_j^\al\,,\quad
  S_i=T_i^\ga+\frac1{2m}(1-(-1)^m),
\]
where the index $\ga$ is fixed but arbitrary. In particular, for $m=2$ we have $T_k^\al=\si_k^\al$
and hence~\footnote{When $m=2$ we take $\ga=1$ in the equation for $S_i$ to agree with the
  standard definition of the Pauli matrices.} $S_i=\si_i^1$,
$S_{ij}=\frac12(1+\bsi_i\cdot\bsi_j)$, where $\si^{\al}$ ($\al=1,2,3$) are the Pauli matrices.
Note that the operators $S_{ij}$ obey the standard permutation algebra
\[
  S_{ij}O_j=O_iS_{ij},\quad S_{ij}O_i=O_jS_{ij},\quad S_{ij}O_k=O_kS_{ij},
\]
where $k\ne i,j$ and $O_k$ is any operator acting on the $k$-th site.

The first class of spin chains we shall be interested in are the Haldane--Shastry (HS),
Polychronakos--Frahm (PF)~\cite{Po93,Fr93,Po94} and Frahm--Inozemtsev (FI)~\cite{FI94} chains.
They can be collectively defined through the formula~\footnote{In what follows sums over Latin
  indices will implicitly range form $1$ to $N$, unless otherwise stated.}
\begin{equation}\label{HANm1}
  H_{\pm} = \sum_{i<j}J_{ij}(1\mp S_{ij})\,,
\end{equation}
where
\begin{align}
  J_{ij}&=\frac{J}{2N^2\sin^2(\xi_i-\xi_j)},\quad\xi_k=\frac{k\pi}N
                                                     \en (\text{HS chain})\,,\label{HS}\\
  J_{ij}&=\frac{J}{N(\xi_i-\xi_j)^2}\,,\quad H_N(\xi_k)=0
                                                \quad (\text{PF chain})\,,\label{PF}\\
  J_{ij}&=\frac{J}{2N^2\sinh^2(\xi_i-\xi_j)}\,,\quad L_N^{c-1}(e^{2\xi_k})=0
                                                        \quad (\text{FI chain})\label{FI}.
\end{align}
Here~$J>0$ is a real constant fixing the energy scale, $H_N$ denotes the Hermite polynomial of
degree $N$ and $L_N^{c-1}$ is a generalized Laguerre polynomial of degree $N$ with positive
parameter $c$. For reasons that will become clear in the sequel, we shall sometimes refer to the
ferromagnetic models $H_+$ as \emph{bosonic} and to the antiferromagnetic ones $H_-$ as
\emph{fermionic}. From the previous expressions we see that the HS chain can be naturally
considered as a circular chain with equally spaced sites and spin-spin interactions inversely
proportional to the square of the chord distance, while the PF and FI chains are better regarded
as linear chains with sites $\xi_k$ defined in Eqs.~\eqref{PF}-\eqref{FI} and respectively
rational (inverse square) or hyperbolic interactions.

The spin chains discussed above are all naturally related to the $A_{N-1}$ classical root system.
The second type of chain we shall deal with is the variant of the HS chain related to the $BC_N$
root system (HS-B chain), whose Hamiltonian shall be taken as~\cite{BPS95,EFGR05}
\begin{subequations}\label{Bchain}
\begin{multline}
  H_\pm=\frac J{4N^2}\sum_{i<j}\bigg(\frac{1\mp S_{ij}}{\sin^{2} (\xi_i-\xi_j)}
  +\frac{1\mp\widetilde{S}_{ij}}{\sin^{2}(\xi_i+\xi_j)}\bigg)\\
  +\frac{J}{8N^2}\sum_i\left(\frac{\be_1}{\sin^{2} \xi_i} + \frac{\be_2}{\cos^2
      \xi_i}\right)\big(1+\vep S_i\big)
\end{multline}
with
\begin{equation}
  P_N^{\be_1-1,\be_2-1}(\cos 2\xi_k)=0.
\end{equation}
\end{subequations}
Here $J>0$, $P_N^{\be_1-1,\be_2-1}$ is a Jacobi polynomial of degree $N$ and parameters
$\be_{1,2}>0$, $\vep=\pm1$, and we have used the
abbreviation~$\tS_{ij}:=S_{ij}S_iS_j=S_iS_jS_{ij}$. This model can be regarded as the \emph{open}
version of the HS chain, with chain sites $z_j:=e^{2\iu\xi_j}$ lying on the upper unit circle
(although in general not uniformly spaced). The spin at $z_j$ interacts both with the remaining
spins at $z_k$ (with $k\ne j$) and their reflections $\bar z_k$ with respect to the real axis, the
interaction strength being inversely proportional to the square of the (chord) distances
$|z_j-z_k|$ and $|z_j-\bar z_k|$.

The HS spin chain was originally introduced as a parent Hamiltonian for the Gutzwiller variational
ground state for the one-dimensional Hubbard model in the limit of infinite on-site energy. In
fact, the (exact) ground state of the HS chain contains a Jastrow factor reminiscent of the
Laughlin wave function in the fractional quantum Hall effect~\cite{La83,Ha91b,AI94}. The
remarkable properties of the HS chain became apparent shortly after its introduction. For
instance, it was soon found that its spinon excitations behave as effective particles obeying
Haldane's fractional statistics~\cite{HHTBP92,Ha91b,Ha93}. This can indeed be regarded as the
simplest realization of anyons in one dimension. The HS chain is completely
integrable~\cite{Ka92,BGHP93,HH93}, and is actually invariant under the Yangian quantum
algebra~$Y(\sla(m))$ even for a finite number of sites, which in part explains the high
degeneracies of its spectrum~\cite{FG15}. It is closely related to the
Wess--Zumino--Novikov--Witten (WZNW) model at level 1~\cite{HHTBP92}, and can also be embedded
into a larger class of models constructed from chiral vertex operators of an appropriate conformal
field theory~\cite{CS10,NCS11}.

A characteristic property of the HS chain that distinguishes it from short-range chains like the
Heisenberg model is the fact that it can be obtained as the strong interaction (large coupling)
limit of an integrable one-dimensional system, namely the (spin) Sutherland
model~\cite{Su71b,Su72,HH92}. In fact, the PF and FI chains can be analogously derived from the
integrable spin Calogero~\cite{Ca71,MP93} and Frahm--Inozemtsev~\cite{In96} dynamical models.
Similarly, the HS-B chain is the large coupling limit of the spin Sutherland model of $BC_N$
type~\cite{Ya95,BPS95,EFGR05}. As first pointed out by Polychronakos~\cite{Po94}, the connection
between the spin chains~\eqref{HANm1}--\eqref{Bchain} and the dynamical spin models of
Calogero--Sutherland type mentioned above can be exploited to derive the chains' partition
functions in closed form~\cite{FG05,EFGR05,BFGR08epl,BFGR10}. From the common structure of these
partition functions and their relation to the representation theory of the Yangian algebra in
terms of border strips and their associated motifs~\cite{KKN97}, a remarkable equivalence between
the $A_{N-1}$ chains~\eqref{HANm1}--\eqref{FI} and certain (inhomogeneous) vertex models was
established in Ref.~\cite{BBH10}. More precisely, the spectrum of the latter chains (with the
correct degeneracy for each energy) is the same as that of a vertex model with $N+1$ vertices
connected by $N$ bonds, each of which can take the values $1,\dots,m$. The energy of a
configuration of this
model, represented by a \emph{bond vector}
\[
  \bsv:=(s_1,\dots,s_N)\in\{1,2,\dots,m\}^N,
\]
can be computed through the formula
\begin{equation}\label{EbsvA}
  E(\bsv)=J\sum_{i=1}^{N-1}\de_\pm(s_i,s_{i+1})\cE(x_i)\,,\quad x_i:=i/N\,.
\end{equation}
Here the \emph{dispersion function} $\cE(x)$, which depends on the chain considered, is given by
\begin{equation}\label{disp}
  \cE(x)=\begin{cases}
    x(1-x),\quad &(\text{HS chain})\\
  x, &(\text{PF chain})\\
  x(x+\ga_N),&(\text{FI chain})
\end{cases}
\end{equation}
with $\ga_N:=(c-1)/N$, while the function $\de_\pm$ (where the $\pm$ sign corresponds to the
double sign in $H_\pm$) is defined by
\begin{equation}\label{delta}
  \de_+(i,j)=\begin{cases}
    0,\en s_i\le s_{i+1}\\
  1,\en s_i> s_{i+1},
\end{cases}
\quad
\de_-(i,j)=\begin{cases}
    0,\en s_i< s_{i+1}\\
    1,\en s_i\ge s_{i+1}.
  \end{cases}
\end{equation}
We thus see that the spin degrees of freedom behave as bosons (resp.~fermions) in the
ferromagnetic (resp.~antiferromagnetic) case. Note also that the vectors with components
$\de_\pm(s_i,s_{i+1})$ are closely connected to the \emph{motifs} introduced by
Haldane~\cite{HHTBP92}. An analogous description of the spectrum of the HS-B chain was recently
found in Ref.~\cite{CFGR20}. More precisely, in this case the vertex model has an additional
vertex and a last bond $s_{N+1}$ assuming the \emph{fixed} (half-integer) value $m_\vep+\frac12$,
where
\[
  m_\vep:=\frac12(m+\vep\pi(m))
\]
and $\pi(m)=\frac12\,\big(1-(-1)^m\big)$ is the parity of $m$. Thus in this case
\begin{equation}\label{EbsvB}
  E(\bsv)=J\sum_{i=1}^{N}\de_\pm(s_i,s_{i+1})\cE(x_i)\,,
\end{equation}
with $\de_\pm$ as above and dispersion function
\begin{equation}
  \label{dispB}
  \cE(x)=x\,\bigg(\ga_N+1-\frac{x}2\bigg),
  \quad\ga_N:=\frac1{2N}(\be_1+\be_2-1)\,.
\end{equation}

The thermodynamics of spin chains of HS type has been studied ever since the early work of
Haldane, who derived an expression for the entropy of the $\su(2)$ HS chain by means of the spinon
description of its spectrum~\cite{Ha91}. Shortly afterwards, Sutherland and Shastry~\cite{SS93}
addressed the general $\su(m)$ case, outlining a complicated procedure for computing the free
energy which involves two successive integrations. This procedure, however, only yields an
explicit expression for $m=2$. Around the same time, a heuristic formula for the free energy per
spin of the PF and FI chains (with no magnetic field or chemical potential term) was presented in
Refs.~\cite{Fr93,FI94}, again only for the $\su(2)$ case. In fact, a systematic study of the
thermodynamics of the $A_{N-1}$ chains~\eqref{HANm1}-\eqref{FI} (with an additional chemical
potential or magnetic field term) using the transfer matrix method was undertaken for the first
time in Ref.~\cite{EFG12}, and extended later to the supersymmetric case in Ref.~\cite{FGLR18}.
The key idea in this respect is the fact that using Eq.~\eqref{EbsvA} (or, actually, its
generalization to allow for a chemical potential term) it is straightforward to express the
partition function as the trace of a product of $N$ site-dependent $m\times m$ transfer matrices.
In our case, it follows from~\eqref{EbsvA} that the partition function of the three $A_{N-1}$
chains can be collectively written as
\[
  Z=\tr\big[A(x_0)A(x_1)\cdots A(x_{N-1})\big]\,,
\]
where the $m\times m$ \emph{transfer matrix} $A(x)$ has entries
\begin{equation}
  \label{Amat}
  A_{\mu\nu}=e^{-\be J\cE(x)\de_\pm(\mu,\nu)}\,,\quad 1\le\mu,\nu\le m.
\end{equation}
Here $\be:=1/T$ is the inverse temperature (taking Boltzmann's constant $k_{\mathrm B}$ as unity),
and the dispersion relation~$\cE(x)$ is given by~\eqref{disp}. Since the matrix $A(x)$ has
positive entries for all $x\in[0,1]$, the classical Perron theorem~\cite{Pe07,GK00} implies that
$A(x)$ has a positive \emph{simple} eigenvalue $\la_1(x)$ which exceeds the modulus of any other
eigenvalue. From this fact it readily follows that in the thermodynamic limit $N\to\infty$ the
free energy per spin $f(T)$ of the chains~\eqref{HANm1}-\eqref{FI} can be expressed as
\begin{equation}
  \label{f}
  f(T)=-T\int_0^1\ln\la_1(x)\,\diff x\,.
\end{equation}
As shown in Ref.~\cite{FG22}, similar expressions are valid for the HS-B chain~\eqref{Bchain}. Indeed,
from~\eqref{EbsvB} we obtain
\[
Z=\tr\big[A(x_1)\cdots A(x_{N-1})B\big],
\]
where $A_{\mu\nu}(x)$ is defined as above but using Eq.~\eqref{dispB} for the dispersion
relation, and the $m\times m$ matrix $B$ has entries
\[
  B_{\mu\nu}=e^{-\be J(\ga_N+1/2)\de_\pm(\mu,m_\vep+\frac12)}\,,\quad 1\le\mu,\nu\le m\,.
\]
Since all the matrices in the expression for the partition function have again positive entries,
it follows from Perron's theorem that the thermodynamic free energy per spin is given by
Eq.~\eqref{f} also in this case (see~\cite{FG22} for details). Note that when computing the
thermodynamic free energy from Eq.~\eqref{f} we must replace the parameter $\ga_N$ in
\eqref{disp}-\eqref{dispB} by
\[
  \ga:=\lim_{N\to\infty}\ga_N\ge0\,.
\]
In fact, Eq.~\eqref{f} has been shown to hold for the $\su(m|n)$ supersymmetric version of the
chains~\eqref{HANm1}--\eqref{Bchain} studied in this paper, even with the addition of a general
chemical potential term~\cite{FGLR18,FG22}. Thus the thermodynamic functions of all of these
models can be computed in closed form provided that the Perron eigenvalue $\la_1(x)$ of the
transfer matrix $A(x)$ in Eq.~\eqref{Amat} can be explicitly found. So far, this has only been
done in the $\su(2)$ case (bosonic or fermionic)~\cite{EFG12,FG22} and in the supersymmetric
case~\footnote{In the truly supersymmetric case the transfer matrix has always a zero eigenvalue,
  which is doubly degenerate for $m=n=2$. This makes it straightforward to diagonalize the latter
  matrix when $1\le m,n\le2$.} with $1\le m,n\le 2$~\cite{FGLR18,FG22}.

The main aim of this paper is to derive a remarkably simple expression for $\la_1(x)$ for all the
HS-type $\su(m)$ chains~\eqref{HANm1}--\eqref{Bchain}, valid for \emph{arbitrary} values of $m$.
Thus the thermodynamic functions of these models can be evaluated in closed form. We stress that
such closed-form expressions had been obtained so far only for the $\su(2)$ case, even at zero
chemical potential.

Another problem we shall address in this work is the study of the critical behavior of the
chains~\eqref{HANm1}--\eqref{Bchain}. As is well known, a strong indication of the critical
character of a model is the low-temperature behavior of its free energy. The reason for this is
that at low temperatures the free energy per unit length of a $(1+1)$-dimensional conformal field
theory (CFT) with central charge $c$ behaves as~\cite{BCN86,Af86}
\begin{equation}\label{fCFT}
f(T)\simeq f(0)-\frac{\pi c T^2}{6 v}
\end{equation}
where $v$ is the Fermi velocity and we are using natural units $\hbar=1$. It is thus expected that
the free energy of a critical system obey the latter asymptotic formula at sufficiently low
temperatures, with $c$ equal to the central charge of the effective CFT governing the model's low
energy behavior. Using the explicit expression~\eqref{f} for the free energy per spin of the
chains~\eqref{HANm1}--\eqref{Bchain}, we shall show that Eq.~\eqref{fCFT} is satisfied for these
models with central charge $c=m-1$ (with the only exception of the FI chain with $\ga=0$). This
result agrees with the calculation in Ref.~\cite{HB00} for the supersymmetric PF chain (using a
different method), and is consistent with the fact that the low energy excitations of the
(original) HS chain~\eqref{HS} are governed by the $\su(m)_1$ WZNW model~\cite{HHTBP92,Sc94,BS96}.
It should be emphasized, however, that~\eqref{fCFT} is only a necessary condition for criticality.
Indeed, a CFT ---and thus a truly critical model--- must have low-energy excitations with a linear
energy-momentum relation and the degeneracy of its ground state should be finite. Using
Eqs.~\eqref{EbsvA}-\eqref{disp} and \eqref{EbsvB}-\eqref{dispB}, we shall prove that both of these
conditions hold in our case. In this way we shall show that the chains~\eqref{HANm1}--\eqref{FI}
(with $\ga>0$ for the FI chain) are critical only in the fermionic (antiferromagnetic) case, which
is again in agreement with the results for the HS chain in Ref.~\cite{BBS08}. On the other hand,
we shall see that the HS-B chain is critical not only in the fermionic case, but also in the
bosonic (ferromagnetic) one when $m=2$, or $m=3$ and $\vep=-1$.

We shall close this section with a few words on the paper's organization. In Section~\ref{sec.f}
we recall the formula for the free energy per spin of $\su(m)$ chains of HS-type in terms of the
Perron eigenvalue of a suitable transfer matrix derived in Refs.~\cite{EFG12} and~\cite{FG22}. We
then evaluate this eigenvalue, thus obtaining a simple closed-form expression for the free energy
of the latter models valid for arbitrary values of $m$. This expression is used in
Section~\ref{sec.thermo} to derive explicit formulas for the energy, entropy and specific heat per
spin. We also study the approximation of these functions by those of an $m$-level system with
uniformly spaced levels, and establish the existence of a single Schottky peak in the specific
heat for all values of $m$. In Section~\ref{sec.crit} we analyze the critical behavior of the
models under study by first determining the low-temperature behavior of their free energy, and
then examining in detail the ground state degeneracy and the existence of low-energy excitations
with linear energy-momentum relation. The paper ends with a technical appendix in which we deduce
the full asymptotic expansion of the free energy per spin at low temperatures used in
Section~\ref{sec.crit}.

\section{Free energy}\label{sec.f}

In this section we shall evaluate the Perron (``dominant'') eigenvalue $\la_1(x)$ of the transfer
matrix~$A(x)$ given by~\eqref{Amat}, which as we have seen determines the free energy per spin of
the HS-type chains~\eqref{HANm1}--\eqref{Bchain} through the integral~\eqref{f}. To this end, we
first recall that according to Perron's theorem the dominant eigenvalue of a positive
matrix~\footnote{A matrix (or, in particular, a vector) is said to be positive if all its entries
  entries are positive.} possesses a positive eigenvector. In fact, a corollary of Perron's
theorem posits that the dominant eigenvalue of a positive matrix is the only eigenvalue possessing
a positive eigenvector. Since this property shall be essential in what follows, we shall briefly
summarize its proof. Indeed, suppose that $\bv$ is a positive eigenvector of a positive matrix $P$
with eigenvalue $\la$, and denote by $\la_1$ the dominant eigenvalue of $P$. Since the transpose
matrix $P^{\mathsf T}$ is also positive and has the same spectrum as $P$, its dominant eigenvalue
is also $\la_1$, and therefore there exists a positive vector~$\bw$ such that
$P^{\mathsf T}\bw=\la_1\bw$. We then have
\[
  \la_1(\bw,\bv)=(P^{\mathsf T}\bw,\bv)=(\bw,P\bv)=\la(\bw,\bv),
\]
where we have used the fact that $P^{\mathsf T}$ and $\la_1$ are real. Since both $\bv$ and $\bw$
are positive vectors, from the latter equality it follows that $\la=\la_1$, as claimed.

In fact, if an eigenvalue~$\la$ of a positive matrix $P$ is known, Perron's theorem provides a
simple test to ascertain whether $\la$ is the dominant eigenvalue of $P$. Indeed, if we denote by
$C^\la_{ij}$ the $(i,j)$ cofactor of the matrix $\la-P$ we then have the elementary identity
\[
  \sum_{j=1}^m(\la\de_{ij}-P_{ij})C^\la_{kj}=\de_{ik}\det(\la-P)=0,
\]
where $m$ is the order of $P$. From the previous identity it follows that \emph{any} row of the
cofactor matrix $(C_{ij}^{\la})_{i,j=1}^m$ is an eigenvector of $P$ with eigenvalue $\la$.
However, as part of the proof of Perron's theorem, it is shown that \emph{all} the cofactors
$C_{ij}^{\la_1}$ corresponding to the dominant eigenvalue $\la_1$ are \emph{positive}~\cite{GK00}.
From the discussion in the previous paragraph we then obtain the following elementary test: an
eigenvalue $\la$ of a positive matrix $P$ is its dominant eigenvalue if and only if \emph{any} row
of the cofactor matrix of $\la-P$ is a positive vector.

Let us now turn to the computation of the dominant eigenvalue of the positive matrix $A(x)$ in
Eq.~\eqref{Amat}. To begin with, from the definition~\eqref{delta} of $\de_+$ it follows that
in the \emph{bosonic} case the matrix $A(x)$ has the following structure
\[
  A(x)= 
  \begin{pmatrix}
    1& 1& \cdots &1 &1\\
    a^m& 1& \cdots &1 &1\\
    \vdots & \vdots &\ddots &\vdots &\vdots \\
    a^m& a^m&\cdots &1& 1\\
    a^m& a^m&\cdots &a^m& 1
  \end{pmatrix},
\]
where we have set~\footnote{For the sake of conciseness, we have omitted the dependence of $a$ on
  $x$.}
\[
  a:=e^{-\be J\cE(x)/m}\,.
\]
An elementary calculation shows that
\[
  \la(x)=\sum_{k=0}^{m-1}a^k=\frac{1-a^m}{1-a}
\]
is an eigenvalue of $A(x)$. We could now apply the previous test to check whether $\la(x)$ is the
dominant eigenvalue of $A(x)$, but in this case it is easier to observe that the vector
\[
\bv=(1,a,\dots,a^{m-1})
\]
is a positive eigenvector of $A(x)$ with eigenvalue $\la(x)$. Indeed,
\begin{align*}
  [A(x)\bv]_k&=\sum_{i=1}^{k-1}a^ma^{i-1}+\sum_{i=k}^ma^{i-1}=
               \sum_{i=k-1}^{m+k-2}a^i\\
  &=a^{k-1}\la(x)=\la(x)v_k\,.
\end{align*}
From the previous discussion it follows that $\la(x)=\la_1(x)$. We thus obtain the following
remarkable formula for the free energy per spin $f_+(T)$ of the \emph{bosonic} HS-type
chains~\eqref{HANm1}--\eqref{Bchain}:
\begin{align}
  f_+(T)&=-T\int_0^1\ln\left(\sum_{k=0}^{m-1}e^{-\be J\cE(x)k/m}\right)\!\diff x\notag\\
        &=
          -T\int_0^1\ln\biggl(\frac{1-e^{-\be J\cE(x)}}{1-e^{-\be J\cE(x)/m}}\biggr)\diff x\notag\\
        &=f_0
       -T\int_0^1\ln\left[
          \frac{\sinh\Bigl(\be J\cE(x)/2\Bigr)}{\sinh\Bigl(\be J
          \cE(x)/(2m)\Bigr)}\right]\!\diff x,
    \label{bsum}
\end{align}
where
\[
  f_0:=\frac J2\bigg(1-\frac1m\bigg)\cE_0\,,\quad\cE_0:=\int_0^1\cE(x)\diff x\,.
\]
The constant~$\cE_0$, which can be easily found from~\eqref{disp}-\eqref{dispB}, takes the values
$1/6$ (HS chain), $1/2$ (PF chain), and $\ga/2+1/3$ (FI and HS-B chains).

An analogous result for the fermionic case can be easily derived noting that
\[
  A_-(x)=A_+(x)+(a^m-1)\mathbb 1\,,
\]
where for clarity's sake we have denoted by $A_-(x)$ (resp.~$A_+(x)$) the matrix $A(x)$ in the
fermionic (resp.~bosonic) case. Hence the dominant eigenvalue of the matrix $A_-(x)$ is given by
\[
  \la_1(x)=\sum_{k=0}^{m-1}a^k+a^m-1=a\sum_{k=0}^{m-1}a^k\,.
\]
Substituting into~\eqref{f} we obtain the following simple formula for the free energy per spin
$f_-(T)$ in the \emph{fermionic} case:
\begin{equation}\label{fsum}
  f_-(T)=f_+(T)+\frac{J\cE_0}m\,.
\end{equation}
For $m=2$, the previous formulas for $f_\pm$ coincide with those in Refs.~\cite{EFG12,FG22}. As an
additional consistency check, note that as $T\to\infty$ from Eqs.~\eqref{bsum}-\eqref{fsum} we
easily obtain
\[
  f(T)\simeq -T\ln m\,,
\]
in agreement with the elementary identity
\[
  Z(T)\underset{T\to\infty}\simeq m^N\,.
\]
\begin{remark}
  The thermodynamic free energy of the PF chain can be computed in closed form in terms of the
  dilogarithm function~\cite{OLBC10}
  \[
    \Li_2(z):=-\int_0^z\frac{\ln(1-t)}t\,\diff t\,,
  \]
  analytic in the cut complex plane $\CC\setminus[1,\infty)$, where the integral is taken along
  any path in the latter set joining the origin to the point $z$. Indeed, we have
  \[
    f_+(T)=\frac{T^2}J\bigg[m \Li_2(e^{-\be J/m})-\Li_2(e^{-\be J})-\frac{\pi^2}6\,(m-1)\bigg].
  \]
\end{remark}
\section{Thermodynamics}\label{sec.thermo}

The energy density $u$, entropy per spin $s$ and specific heat per spin $\cV$ of the HS-type
chains~\eqref{HANm1}--\eqref{Bchain} can be readily computed using
Eqs.~\eqref{bsum}-\eqref{fsum}. Before doing so, to simplify our formulas we shall set
without loss of generality $J=1$, so that energy and temperature become dimensionless. With this
proviso, we find
\begin{widetext}
\begin{align}
  u_\pm&=\frac{\pd}{\pd\be}(\be f_\pm)=\frac12\bigg(1\mp\frac1m\bigg)\cE_0
     -\frac12\int_0^1\cE(x)\left[
     \coth\biggl(\frac{\be\cE(x)}2\biggr)
     -\frac1m\coth\biggl(\frac{\be\cE(x)}{2m}\biggr)
     \right]\!\diff x,
          \label{u}\\
  s&=\be(u_\pm-f_\pm)=\int_0^1\left\{\ln\left[\frac{\sinh\left(\frac{\be\cE(x)}2\right)}%
    {\sinh\left(\frac{\be\cE(x)}{2m}\right)}\right]
    -\frac{\be\cE(x)}2\bigg[
    \coth\biggl(\frac{\be\cE(x)}2\biggr)
-\frac1m\coth\biggl(\frac{\be\cE(x)}{2m}\biggr)\bigg]
    \right\}\!\diff x,
     \label{s}\\
  \cV&=-\be^2\frac{\pd u_\pm}{\pd\be}=\frac{\be^2}4\int_0^1\cE(x)^2\left[
       \frac1{m^2}\csch^2\biggl(\frac{\be\cE(x)}{2m}\biggr)-
       \csch^2\biggl(\frac{\be\cE(x)}{2}\biggr)\right]\!\diff x.
     \label{cV}
\end{align}
\end{widetext}
Using the explicit formula for the free energy per spin of the PF chain from the previous section
we readily find the following simple expressions for its thermodynamic functions:
\begin{align*}
  u_+&=u_--\frac1{2m}=-f_+-T\ln\left(\frac{1-e^{-\be}}{1-e^{-\be/m}}\right),\\
  s&=-2\be f_+-\ln\left(\frac{1-e^{-\be}}{1-e^{-\be/m}}\right),\\
  \cV&=-2\be f_+-2\ln\left(\frac{1-e^{-\be}}{1-e^{-\be/m}}\right)
       +\be\left(\frac{1}{e^\be-1}-\frac{1/m}{e^{\be/m}-1}\right).
\end{align*}

\begin{figure}[h]
  \includegraphics[height=.3\columnwidth]{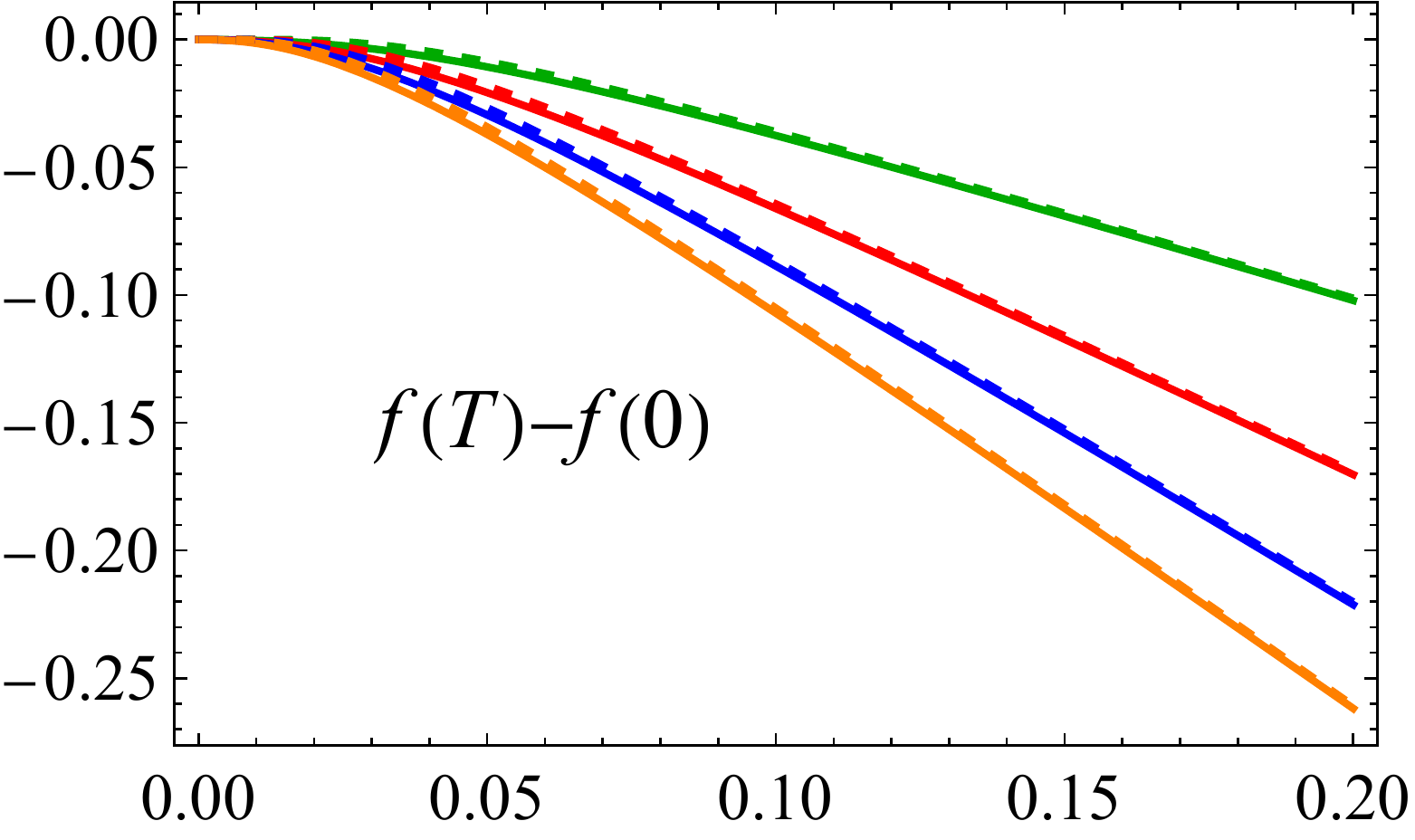}\hfill
  \includegraphics[height=.3\columnwidth]{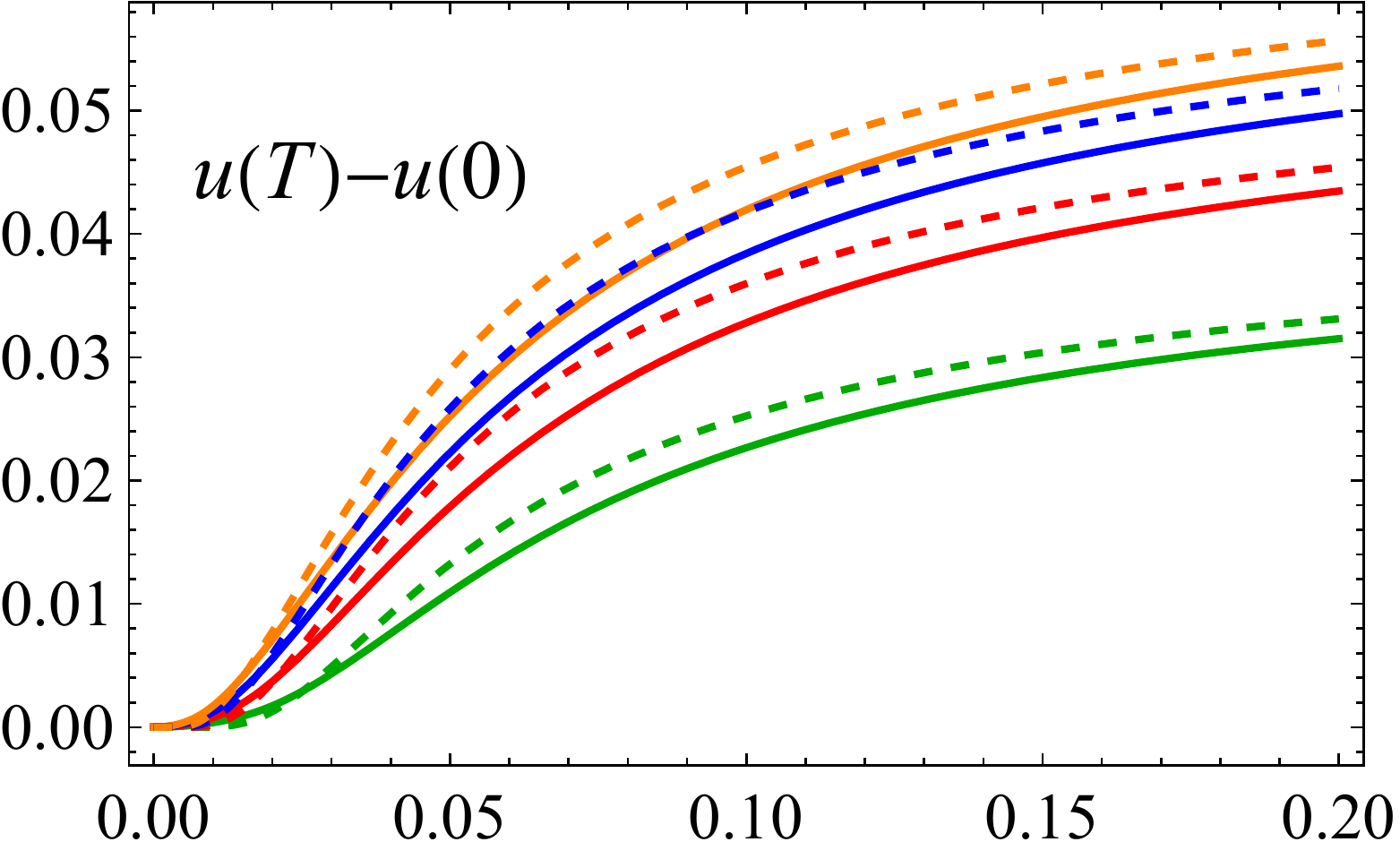}\\[3pt]
  \null\en\kern3pt\includegraphics[height=.3\columnwidth]{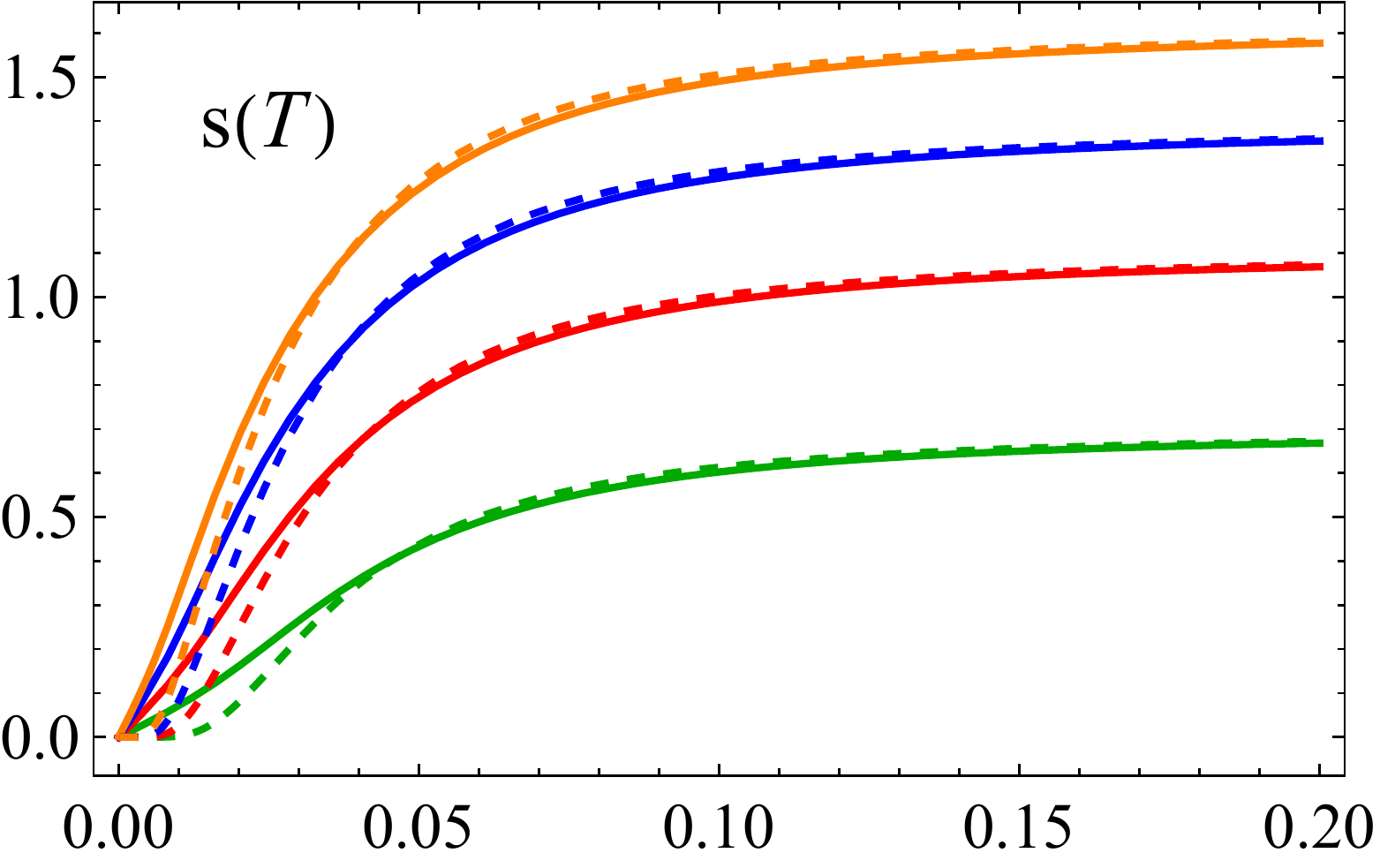}\hfill
  \includegraphics[height=.3\columnwidth]{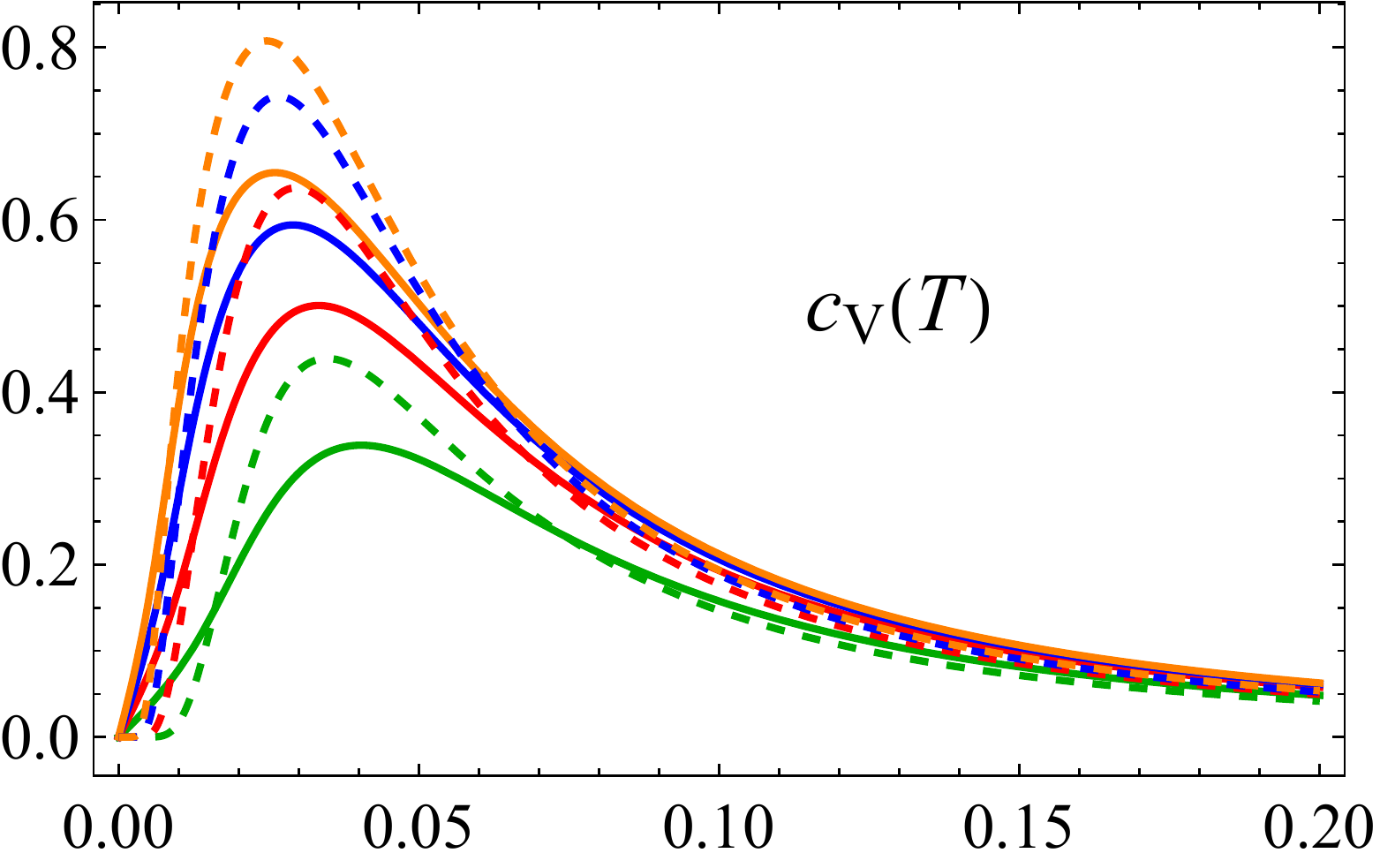}
  \caption{Thermodynamic free energy, energy, entropy, and specific heat per spin of the $\su(m)$
    HS chain~\eqref{HANm1}-\eqref{HS} with $m=2,3,4,5$ (green, red, blue, and orange solid lines,
    respectively), compared to their counterparts for an $m$-level system (dashed lines).}
  \label{fig.thermo}
\end{figure}

From the first equality in Eq.~\eqref{bsum} it follows that for large $T$ (i.e., for $T\gg\cE(1)$
for the PF, FI and HS-B chains or $T\gg\cE(1/2)=1/4$ for the HS chain) the ferromagnetic free
energy per spin $f_+(T)$ can be well approximated replacing $\cE(x)$ by its mean value~$\cE_0$
over the interval $[0,1]$. In other words, we have
\[
  f_+(T)\underset{T\to\infty}\simeq
  -T\ln\left(\sum_{k=0}^{m-1}e^{-k\be\frac{\cE_0}m}\right)=:f_m(T).
\]
The right-hand side is the partition function of an $m$-level system with uniformly spaced levels
$E_k=k\cE_0/m$, $k=0,\dots,m-1$. In fact, at sufficiently high temperatures the thermodynamic
functions of the HS-type chains~\eqref{HANm1}-\eqref{Bchain} behave qualitatively as those of the
corresponding $m$ level system, as can be seen, for instance, from Fig.~\ref{fig.thermo} for the
HS chain with $m=2,\dots,5$ (we omit the corresponding plots for the PF, FI and HS-B chains, which
are very similar). It should be noted, however, that at low temperatures the thermodynamic
functions of the HS-type chains behave quite differently than those of their corresponding
$m$-level system. Indeed, at low temperatures the free energy of the $m$-level system,
\begin{align*}
  f_m(T)&=-T\ln\left(\frac{1-e^{-\be\cE_0}}{1-e^{-\be\cE_0/m}}\right)\\
  &=-Te^{-\be\cE_0/m}+O(Te^{-2\be\cE_0/m}),
\end{align*}
is exponentially small, and so are its remaining thermodynamic functions. On the other hand, from
the discussion in the next section (cf.~Eq.~\eqref{fasympfin}) it follows that for the HS-type
chains $f_\pm(T)-f_\pm(0)\sim -T^2$ as $T\to0+$ (or $\sim -T^{3/2}$ for the FI chain with $\ga=0$;
cf.~Eq.~\eqref{fFI0}). Thus $u-u(0)\sim T^2$, $s\sim T$ and $c_V\sim T$ as $T\to0+$ (or
$u-u(0)\sim T^{3/2}$, $s\sim T^{1/2}$ and $c_V\sim T^{1/2}$ for the FI chain with $\ga=0$).

\begin{figure}[h]
  \includegraphics[width=.49\columnwidth]{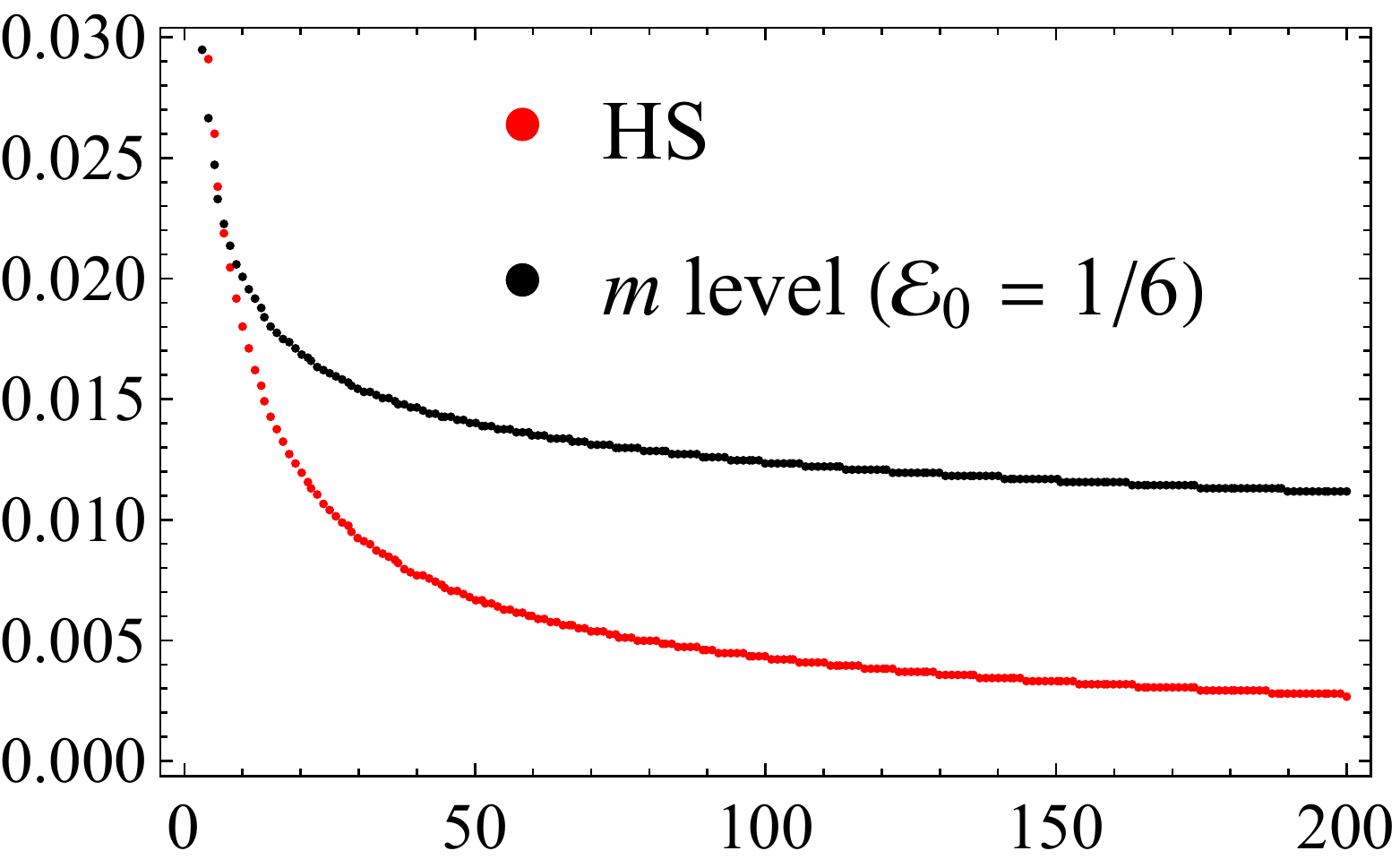}\hfill
  \includegraphics[width=.49\columnwidth]{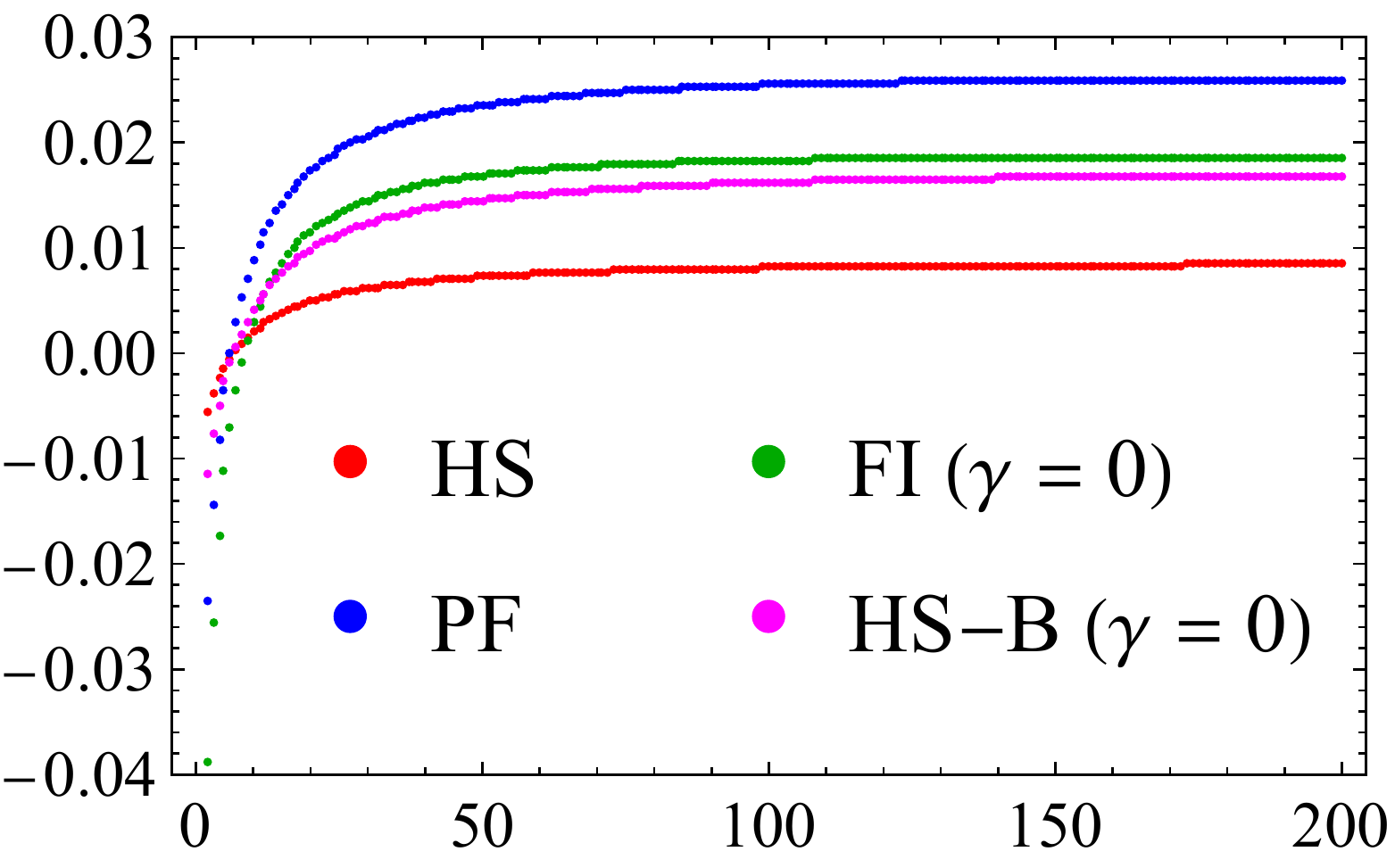}
  \caption{Left: Temperature $T_m$ of the Schottky peak of the $\su(m)$ HS chain
    with~$m=2,\dots,200$ compared to its rough $m$-level approximation~$\cE_0/(2t_m)=1/(12t_m)$.
    Right: difference $\cE_0/(2t_m)-T_m$ for the HS, PF, FI HS-B chains (with $\ga=0$ for the last
    two chains) for $m=2,\dots,200$.}
  \label{fig.Schottky}
\end{figure}

From Fig.~\ref{fig.thermo} (and similar plots for the PF, FI and HS-B chains that we are not
displaying) it is apparent that, at least for low values of $m$, the specific heat per spin of the
HS-type chains features a single absolute maximum (i.e., a Schottky peak). This behavior can be
qualitatively understood by analyzing the specific heat of the corresponding $m$-level system,
given by
\[
  c_{\mathrm V,m}=t^2\left[\frac1{m^2}\csch^2(t/m)-\csch^2t\right],\quad
  t:=\frac{\be\cE_0}2.
\]
Indeed, although we have not found an analytic proof, we have compelling numerical evidence that
$c_{\mathrm V,m}(t)$ has a single critical point~$t_m$ in the half-line $(0,\infty)$ which is an
absolute maximum. It is therefore to be expected that the specific heat of the $\su(m)$ HS-type
chains feature a single Schottky peak at a temperature $T_m$ of the order of $\cE_0/(2t_m)$. In
fact, for low values of $m\lesssim10$ the Schottky temperature $T_m$ is reasonably well
approximated by the $m$-level value $\cE_0/(2t_m)$, as can be seen from Fig.~\ref{fig.Schottky}
(left) for the HS chain (and similarly for the other HS-type chains). On the other hand, for large
$m$ the difference $\cE_0/(2t_m)-T_m$ appears to tend to a constant positive value, while
$T_m\to0$ (see Fig.~\ref{fig.Schottky}, right).

\section{Critical behavior}\label{sec.crit}

As mentioned in the Introduction, one of the hallmarks of criticality is the low-temperature
behavior of the free energy, given by Eq.~\eqref{fCFT}. In our case, using
Eqs.~\eqref{bsum}-\eqref{fsum} with $J=1$ we obtain
\begin{align}
  f_\pm(T)-f_\pm(0)=&-T\int_0^1\ln\left(1-e^{-\be\cE(x)}\right)\diff x\notag\\
                    &+T\int_0^1\ln\left(1-e^{-\be\cE(x)/m}\right)\diff x\,.
                      \label{fpm}
\end{align}
As $T\to0+$, the main contribution to both of these integrals comes from a small neighborhood of
the points where the dispersion relation $\cE(x)$ vanishes, i.e., $x=0,1$ for the HS chain and
$x=0$ for the PF, FI and HS-B chains. For the latter three chains $\cE(x)$ is monotonically
increasing over the interval $[0,1]$, so that performing the changes of variables $y=\be\cE(x)$
and $y=\be\cE(x)/m$ in the integrals in~\eqref{fpm} we have
\begin{multline}
  f_\pm(T)-f_\pm(0)=-T^2\int_0^{\be\cE(1)}\ln\left(1-e^{-y}\right)(\cE^{-1})'(Ty)\,\diff y\\
  +mT^2\int_0^{\frac{\be\cE(1)}m}\ln\left(1-e^{-y}\right)(\cE^{-1})'(mTy)\,\diff y,
  \label{gdef}
\end{multline}
where $\cE^{-1}:0\to\cE(1)$ denotes the inverse function of $\cE:0\to1$. Since the main
contribution to both integrals comes from the point $y=0$, we can approximate $(\cE^{-1})'(Ty)$
and $(\cE^{-1})'(mTy)$ by $(\cE^{-1})'(0)=1/\cE'(0)$ (assuming that $\ga>0$ for the FI chain) and
push the upper limit in each integral to $+\infty$, thus obtaining
\begin{align}
  f_\pm(T)-f_\pm(0)&=\frac{(m-1)T^2}{\cE'(0)}\int_0^\infty\ln\left(1-e^{-y}\right)\,\diff y+O(T^3)
                     \notag\\
                   &=-\frac{(m-1)T^2}{\cE'(0)}\,\Li_2(1)+O(T^3)\notag\\
                   &=-\frac{(m-1)\pi^2}{6\cE'(0)}\,T^2+O(T^3);
                     \label{fasymp}
\end{align}
see the appendix for more details~\footnote{In fact, it is shown in the appendix that for the PF
  chain the $O(T^3)$ term is actually $O(T^2e^{-\be/m})$.}. It follows from~\eqref{fasymp} that the
free energy per spin of the PF, FI (with $\ga>0$) and HS-B chains behaves as that of a CFT with
central charge $c=m-1$. To see this, note first of all that the variable $x$ can be regarded as
$p/\pi$, where $p$ is the momentum (defined modulo $2\pi$). Indeed, since the dispersion relation
is monotonic the interval $0\le x\le 1$ corresponds to the positive momentum range $0\le p\le\pi$
and not to the full range $-\pi\le p\le\pi$. As the relation between energy $\cE(x)$ and momentum
$p=\pi x$ is linear near $p=0$, the Fermi velocity is given by
\begin{equation}\label{vPFFIB}
  v=\left.\frac{\diff\cE}{\diff p}\right|_{p=0}=\frac{\cE'(0)}\pi\,,
\end{equation}
so that~\eqref{fasymp} can indeed be written as
\begin{equation}
  \label{fasympfin}
  f_\pm(T)-f_\pm(0)=-\frac{(m-1)\pi}{6v}\,T^2+O(T^3).
\end{equation}

For the HS chain~\eqref{HANm1}-\eqref{HS}, the dispersion
relation~$\cE(x)$ is symmetric about $x=1/2$ and increasing in the interval $[0,1/2]$. Hence we
can write
\begin{multline}
  f_\pm(T)-f_\pm(0)=-2T\int_0^{1/2}\ln\left(1-e^{-\be\cE(x)}\right)\diff x\\
                    +2T\int_0^{1/2}\ln\left(1-e^{-\be\cE(x)/m}\right)\diff x\,,
                      \label{fpmHS}
\end{multline}
and proceeding as above we arrive at
\begin{equation}\label{fasympHS}
  f_\pm(T)-f_\pm(0)=-\frac{(m-1)\pi^2}{3\cE'(0)}\,T^2+O(T^3)
\end{equation}
(see again the appendix for details on the error term). However, in this case the symmetry of the
dispersion relation about $x=1/2$ implies that the relation between $x$ and $p$ is $p=2\pi x$ (the
interval $0\le x\le 1$ now corresponds to the full momentum range $0\le p\le 2\pi$). Hence
\begin{equation}\label{vHS}
  v=\left.\frac{\diff\cE}{\diff p}\right|_{p=0}=\frac{\cE'(0)}{2\pi}\,,
\end{equation}
and~\eqref{fasympfin} also holds in this case. Thus Eq.~\eqref{fasympfin} is valid for all the
HS-type chains, except for the FI chain with $\ga=0$. (In fact, as shown in the appendix, for the
latter chain $f_\pm(T)-f_\pm(0)\sim -T^{3/2}$ as $T\to0+$.)

As mentioned in the Introduction, to ascertain the criticality of a quantum system we must also
examine the degeneracy of its ground state and study its low-energy excitations. Both of these
problems can be addressed with the help of Eqs.~\eqref{EbsvA}-\eqref{disp} and
\eqref{EbsvB}-\eqref{dispB} for the energy spectrum. We shall start by determining the ground
state of each of the chains~\eqref{HANm1}-\eqref{Bchain} and its degeneracy.

Consider, first, the PF, FI and HS-B chains, whose dispersion relation~$\cE(x)$ is monotonically
increasing over the interval $[0,1]$. In the fermionic case $\de(s_i,s_i+1)=0$ if and only if
$s_i<s_{i+1}$, and hence the bond vectors $\bsv_g$ yielding the ground state of the PF and FI
chains are obtained by placing~\footnote{We use the notation $\lfloor x\rfloor$ to denote the
  integer part of the real number $x$.} $r:=\left\lfloor N/m\right\rfloor$ sequences $(1,\dots,m)$
starting from the right end and filling the cremaining $N-rm$ components with an increasing
sequence $(s_1,\dots,s_{N-rm})\in\{1,\dots,m\}^{N-rm}$:
\begin{equation}
\begin{aligned}
  &\bsv_g=(s_1,\dots,s_{N-rm},1,\dots,m,\dots,1,\dots,m),\\
  &1< s_1<\cdots<s_{N-rm}\,.
\end{aligned}\label{bvPF-FI}
\end{equation}
The ground state degeneracy is therefore $\binom{m}{N-rm}\ll N$ as $N\to\infty$. In the case of
the HS-B chain, since $\de(s_N,s_{N+1})=\de(s_N,m_\vep+\frac12)$, to obtain the ground state we
must take $s_N=m_\vep$. Hence the ground state bond vectors are in this case
\begin{align*}
  &\bsv_g=(s_1,\dots,s_{N-rm},1,\dots,m,\dots,1,\dots,m,1,\dots,m_\vep),\\
  &1< s_1<\cdots<s_{N-rm},
\end{align*}
where now $r=\left\lfloor(N-m_\vep)/m\right\rfloor$. The ground state degeneracy is again
$\binom{m}{N-rm}\ll N$. Thus the degeneracy of the fermionic PF, FI and HS-B chains remains finite
in the thermodynamic limit. The situation is completely different in the bosonic case, since now
$\de(s,s)=0$. Thus the ground state bond vectors of the PF and FI chains are of the form
\begin{equation}\label{bvbosonic}
  \bsv_g=(\underbrace{s_1,\dots,s_1}_{k_1},\dots,\underbrace{s_r,\dots,s_r}_{k_r})
\end{equation}
with $k_1+\cdots+k_r=N$ and $1\le s_1<\cdots<s_r\le m$. Thus in this case the ground state
degeneracy is given by
\[
  d_g=\sum_{r=1}^m\cP(N;r)\binom mr,
\]
where $\cP(N;r)$ denotes the number of partitions of the integer $N$ in $r$ parts. In particular,
the ground state degeneracy clearly tends to infinity in the thermodynamic limit. On the other
hand, for the bosonic HS-B chain we must impose the additional restriction $s_r\le m_\vep$, so
that in this case we have
\[
d_g=\sum_{r=1}^{m_\vep}\cP(N;r)\binom{m_\vep}r.
\]
It follows that in this case the ground state is non-degenerate if and only if $m_\vep=1$, i.e.,
for $m=2$ or $m=3$ and $\vep=-1$. The ground state degeneracy is otherwise infinite (at least
$N +1$) in the thermodynamic limit.

The above analysis must be slightly modified in the case of the HS chain, whose dispersion
relation is increasing over the interval $[0,1/2]$ and symmetric about $1/2$. For this reason, to
the bond vectors~\eqref{bvPF-FI} we should add their ``reflected'' counterparts
\[
(1,\dots,m,\dots,1,\dots,m,s_1,\dots,s_{N-rm}).
\]
Thus in this case the ground state degeneracy is again finite in the thermodynamic limit
($2\binom{m}{N-rm}$ when $N$ is not a multiple of $m$, or $1$ when it is). Finally, in the bosonic
case the ground state bond vector is still of the form~\eqref{bvbosonic}, and hence the ground
state degeneracy is infinite in the thermodynamic limit. In summary, from the analysis of the
ground state degeneracy we conclude that only the fermionic chains~\eqref{HANm1}-\eqref{Bchain},
and the bosonic HS-B chain with $m=2$ or $m=3,\vep=-1$, can be truly critical. It is interesting
to observe in this respect that the bosonic $\su(3)$ HS-B chain with $\vep=1$ has exactly the same
thermodynamic functions as its counterpart with $\vep=-1$, but only the latter can be critical.

Let us now turn to the study of the low energy excitations over the ground state in the possible
critical cases identified in the previous paragraph. To begin with, consider the fermionic
$\su(m)$ HS chain. A low energy excitation over a ground state with bond vector~$\bsv_g$ is
obtained, for instance, replacing a component $s_i$ with $i\ll N$ (or $N-i\ll N$) and
$s_i<s_{i+1}$, by $s'_i\ge s_{i+1}$. In the first case we add an energy
\[
  \De E=\cE(x_i)\simeq\cE'(0)x_i=O(1/N),
\]
while in the second one
\[
  \De E=\cE(x_i)=\cE(1-x_i)\simeq\cE'(0)(1-x_i)=O(1/N).
\]
On the other hand, it is well known~\cite{HHTBP92} that the state described by a bond
vector~$\bsv$ has momentum
\[
  P(\bsv)=2\pi\sum_{i=1}^{N-1}x_i\de(s_i,s_{i+1})\quad\mod{2\pi}.
\]
Hence by replacing $s_i$ by $s_i'$ we add a momentum
\[
  \De P = \pm2\pi x_i\quad\mod{2\pi}\,
\]
where the ``$+$'' sign corresponds to the case $i\ll N$. Thus in this case the model is critical,
and its Fermi velocity is given by
\[
  v=\lim_{N\to\infty}\left|\frac{\De E}{\De P}\right|=\frac{\cE'(0)}{2\pi},
\]
in agreement with Eq.~\eqref{vHS}. Note that this conclusion is consistent with the discussion in
Ref.~\cite{BBS08} for the $\su(m|n)$-supersymmetric HS chains.

For the fermionic PF, FI and HS-B chains the situation is quite different, as the Hamiltonian of
these models is not invariant under translations along the lattice and thus momentum is not
conserved. Of course, these models still possess low energy excitations obtained by exciting a
component of the ground state bond vector with index $i\ll N$ (both in the fermionic case and for
the bosonic HS-B chain with $m=2$ or $m=3,\vep=-1$), with energy
\[
  \De E=\cE(x_i)\simeq\cE'(0)x_i=O(1/N).
\]
By analogy with the HS chain, we assign to an energy eigenstate with bond vector~$\bsv$ an
effective momentum
\begin{equation}\label{PPFFIB}
  P(\bsv)=\pi\sum_{i=1}^{N-\eta}x_i\de(s_i,s_{i+1})\quad\mod{2\pi},
\end{equation}
where $\eta=0$ for the HS-B chain and $\eta=1$ otherwise. Note that, as explained above, in this
case the factor multiplying the sum is $\pi$ instead of $2\pi$, since the dispersion relation of
the PF, FI, and HS-B chains is monotonically increasing over the interval $[0,1]$, and thus this
interval represents only the positive momentum range. With this definition the change in effective
momentum of the low energy excitations described above is $\De P=\pi x_i$, and hence their Fermi
velocity is given by Eq.~\eqref{vPFFIB}. In summary, the HS-type chains
chains~\eqref{HANm1}-\eqref{Bchain} are all critical in the fermionic case, while in the bosonic
case only the HS-B chain is critical when $m=2$ or $m=3,\vep=-1$.
\begin{remark}
  The PF chain is known to possess Yangian invariance~\cite{HI95}, and it is conjectured that the
  same is true for the FI chain in view of the structure of its partition function and the high
  degeneracy of its spectrum~\cite{BFGR10}. Likewise, the HS-B chain is also known to have twisted
  Yangian symmetry~\cite{BPS95}. It is thus conceivable that the effective momentum~\eqref{PPFFIB}
  could be related to the eigenvalues of one of the conserved Yangian generators for these models.
\end{remark}

\section{Conclusions}\label{sec.conc}

In this paper we have completely determined in closed form the thermodynamics of the $\su(m)$ spin
chains of Haldane--Shastry type (with zero chemical potential) \eqref{HANm1}-\eqref{Bchain} for
all $m$. Our method relies on the fact that the energy spectrum of these models coincides with
that of an appropriate vertex model, which makes it possible to express their thermodynamic free
energy as an integral involving the Perron eigenvalue of a position-dependent $m\times m$ transfer
matrix. We have been able to compute this eigenvalue in closed form for arbitrary values of $m$ by
applying the classical Perron theorem on positive matrices and some of its consequences. This
yields an explicit expression for the free energy per spin of these models, which fully determines
their thermodynamics. We have found that at sufficiently high temperatures the thermodynamic
functions of the $\su(m)$ HS-type chains behave qualitatively as those of an $m$-level system with
uniformly spaced levels. In particular, for all values of $m$ the specific heat features a single
Schottky peak, whose temperature is close to that of the corresponding $m$-level system for
$m\lesssim 10$.

Using our explicit formula for the free energy per spin, we have also examined the critical
behavior of the $\su(m)$ HS-type chains~\eqref{HANm1}-\eqref{Bchain}. We have first shown that the
low-temperature behavior of the free energy coincides with that of a CFT with central charge
$c=m-1$, both in the ferromagnetic (bosonic) and the antiferromagnetic (fermionic) regimes (with
$\ga>0$ for the FI chain). This low temperature behavior of the free energy is, however, only a
necessary condition for criticality. To ascertain whether the models under study are critical or
not, we have used the motif-based description of their spectrum to study the degeneracy of their
ground state and the existence of low-energy excitations with a linear energy-momentum relation.
In this way we have shown that the antiferromagnetic chains are all critical (again, with $\ga>0$
for the FI chain), whereas in the ferromagnetic case only the $\su(2)$ and $\su(3)$ HS-B chain
(with $\vep=-1$ in the latter case) are critical.

Our main result does not appear to be easily generalizable to the case of non-zero chemical
potential, and in fact the known expressions of the thermodynamic functions in the latter case for
low values of $m$ have a more complicated structure than those found in this paper. Note, however,
that our closed-form expression for the Perron eigenvalue in the zero chemical potential case
could be used to identify and approximate this eigenvalue for sufficiently small values of the
chemical potentials, which would in turn yield a corresponding approximation for the thermodynamic
functions.

\section*{Acknowledgments}
This work was partially supported by grants PGC2018-094898-B-I00 from Spain's Mi\-nis\-te\-rio de
Ciencia, Innovaci\'on y Universidades and~G/6400100/3000 from Universidad Complutense de Madrid.

\appendix
\section{Asymptotic expansion of the free energy}

In this appendix we shall provide the details of the computation of the asymptotic approximation
to the free energy per spin of the chains~\eqref{HANm1}--\eqref{Bchain} at low temperatures (see,
e.g., Eq.~\eqref{fasymp}). In fact, we shall derive a complete asymptotic expansion of the latter
function for $T\to0+$.

Consider, to begin with, the FI chain with $\ga=0$, for which $\cE(x)=x^2$ and hence
$\cE^{-1}(y)=\sqrt y$, $0\le y\le 1$. From Eq.~\eqref{gdef} we have
\begin{equation}\label{gfasymp}
  f_\pm(T)-f_\pm(0)=-T^2\big(g(T)-mg(mT)\big)
\end{equation}
with
\[
  g(T)=\frac12\,T^{-1/2}\int_0^\be\ln(1-e^{-y})\,\frac{\diff y}{\sqrt y}\,.
\]
Since
\begin{multline*}
  \left|\int_0^\infty\ln(1-e^{-y})\,\frac{\diff y}{\sqrt y}
    -\int_0^\be\ln(1-e^{-y})\,\frac{\diff y}{\sqrt y}\right|\\
  =-\int_\be^\infty\ln(1-e^{-y})\,\frac{\diff y}{\sqrt y}
    =O(T^{1/2}e^{-\be})
\end{multline*}
we can write
\begin{align*}
  g(T)&=\frac12\,T^{-1/2}\int_0^\infty\ln(1-e^{-y})\,\frac{\diff y}{\sqrt y}
        +O(e^{-\be})\\
  &=-\frac{\sqrt\pi}2\,\ze(3/2)T^{-1/2}+O(e^{-\be}),
\end{align*}
where $\ze(z):=\sum_{n\ge1}n^{-z}$ is Riemann's zeta function. From Eq.~\eqref{gfasymp} we finally
obtain
\begin{multline}\label{fFI0}
  f_\pm(T)-f_\pm(0)\\=-\left(\sqrt m-1\right)\frac{\sqrt\pi}2\,\ze(3/2)T^{3/2}+O(T^2e^{-\be/m}),
\end{multline}
which shows that the FI chain is not critical for $\ga=0$.

Consider next the PF chain. In this case we simply have
\[
  \int_0^1\ln(1-e^{-\be x})\,\diff x=T\int_0^\be(1-e^{-y})\,\diff y =-\frac{\pi^2T}6+O(Te^{-\be}),
\]
and from~\eqref{fpm} we find
\[
  f_\pm(T)-f_\pm(0)=-(m-1)\frac{\pi^2T^2}6+O(T^2e^{-\be/m}).
\]
This is in agreement with Eq.~\eqref{fasymp}, since in this case $\cE(x)=x$.

Let us next deal with the FI and HS-B chains with $\ga>0$, for which we respectively have
\[
  (\cE^{-1})'(z)=
  \begin{cases}
    (\ga^2+4z)^{-1/2},& 0\le z\le\ga+1\\[3pt]
    \left[(\ga+1)^2-2z)\right]^{-1/2},&0\le z\le\ga+\frac12.
  \end{cases}
\]
Since in both cases all the derivatives of $\cE(z)$ are bounded in their respective domains we
have
\[
  (\cE^{-1})'(z)=\sum_{k=0}^nc_k\frac{z^k}{k!}+O(z^{n+1})\,,\quad
  c_k:=\left.\frac{\diff^{k+1}(\cE^{-1})}{\diff z^{k+1}}\right|_{z=0}.
\]
From these formulas we obtain the estimates
\begin{align*}
  g(T)&=\int_0^{\be\cE(1)}\ln\left(1-e^{-y}\right)(\cE^{-1})'(Ty)\,\diff y\\
  &=\sum_{k=0}^nc_kT^k\int_0^{\be\cE(1)}\frac{y^k}{k!}\,\ln(1-e^{-y})\,\diff y
  +O(T^{n+1}).
\end{align*}
Moreover, since
\[
  \int_{\be\cE(1)}^\infty\frac{y^k}{k!}\,\ln(1-e^{-y})\,\diff y
  =O(\be^ke^{-\be\cE(1)}).
\]
the upper limit in all of the integrals above can be pushed to infinity at the cost of an
exponentially small term. We thus have
\begin{align*}
  g(T)&=\sum_{k=0}^nc_kT^k\int_0^{\infty}\frac{y^k}{k!}\,\ln(1-e^{-y})\,\diff y
        +O(T^{n+1})\\
  &=-\sum_{k=0}^nc_k\ze(k+2)T^k+O(T^{n+1})\,,
\end{align*}
which is equivalent to the infinite asymptotic expansion
\begin{equation}\label{gasymp}
  g(T)\sim-\sum_{k=0}^\infty c_k\ze(k+2)T^k.
\end{equation}
By Eq.~\eqref{gfasymp}, the latter formula yields the following asymptotic series for the free
energy per spin of the FI and HS-B chains with $\ga>0$:
\begin{equation}
  \label{FIHS-Bas}
  f_\pm(T)-f_\pm(0)\sim-\sum_{k=0}^\infty(m^{k+1}-1)c_k\ze(k+2)T^{k+2}.
\end{equation}
The coefficients $c_k$ can be easily computed in both cases, with the result
\[
  c_k=  
  \begin{cases}
    (-2)^k(2k-1)!!\,\ga^{-(2k+1)}&(\text{FI})\\[6pt]
    (-1)^k(2k-1)!!\,(\ga+1)^{-(2k+1)}&(\text{HS-B}),
  \end{cases}
\]
where $(-1)!!:=1$. In particular, from Eqs.~\eqref{disp}-\eqref{dispB} it easily follows that the
first term in the asymptotic series~\eqref{FIHS-Bas} coincides with Eq.~\eqref{fasymp}.

The above argument must be slightly modified to deal with the HS chain and the HS-B chain with
$\ga=0$, since in both cases $(\cE^{-1})'$ becomes infinite at the right endpoint~$\cE(1)$. For
instance, for the HS chain we have
\[
  (\cE^{-1})'(z)=(1-4z)^{-1/2},\quad 0\le z\le 1/4;
\]
note that in this case $\cE^{-1}$ is the inverse of $\cE:[0,1/2]\to[0,1/4]$, since $\cE(x)$ is
increasing over $[0,1/2]$ and symmetric about the half-point $x=1/2$. Using Eq.~\eqref{fpmHS} and
performing the change of variable $\be\cE{x}=y$ we again arrive at Eq.~\eqref{gfasymp}, where
$g(T)$ is now given by
\begin{equation}
  g(T)=2\int_0^{\be/4}\ln\left(1-e^{-y}\right)(1-4Ty)^{-1/2}\,\diff y.
  \label{gHS}
\end{equation}
To deal with the divergence of the last term at the upper limit of the integral, we first note that
\[
  (1-4z)^{-1/2}=\frac{(-1)^{n+1}}{2^{n+1}(2n+1)!!}\,\frac{\diff^{n+1}}{\diff
      z^{n+1}}(1-4z)^{\frac{2n+1}2}\,.
\]
We next define
\[
  h(z)=\frac{(-1)^{n+1}}{2^{n+1}(2n+1)!!}\left[(1-4z)^{\frac{2n+1}2}
    -\sum_{k=0}^{2n+1}c_k\frac{z^k}{k!}\right],
\]
with
\[
  c_k=\left.\frac{\diff^k}{\diff z^k}\right|_{z=0}(1-4z)^{\frac{2n+1}2},
\]
so that $h(z)=O(z^{2n+2})$. Differentiating $n+1$ times the expression for $h(z)$ we arrive at the
identity
\begin{align*}
  &(1-4z)^{-1/2}\\
  &=h^{(n+1)}(z)+\frac{(-1)^{n+1}}{2^{n+1}(2n+1)!!} \frac{\diff^{n+1}}{\diff
    z^{n+1}}\sum_{k=n+1}^{2n+1}c_k\frac{z^k}{k!}\\
  &=h^{(n+1)}(z)+\frac{(-1)^{n+1}}{2^{n+1}(2n+1)!!}\sum_{j=0}^{n}c_{n+j+1}\frac{z^j}{j!}\,.
\end{align*}
Taking into account that
\[
  c_{n+j+1}=(2n+1)!!(2j-1)!!(-1)^{n+1}2^{n+j+1}
\]
we finally obtain
\[
(1-4z)^{-1/2}=h^{(n+1)}(z)+\sum_{j=0}^n2^j(2j-1)!!\,\frac{z^j}{j!}.
\]
Substituting this expression into Eq.~\eqref{gHS} yields
\begin{align*}
  g(T)=&\sum_{j=0}^{n}2^{j+1}(2j-1)!!\,T^j\int_0^{\be/4}\frac{y^j}{j!}\ln(1-e^{-y})\diff y\\
  &+2\be^{n+1}\int_0^{\be/4}\ln(1-e^{-y})\frac{\diff^{n+1}}{\diff y^{n+1}}h(Ty)\diff y.
\end{align*}
The asymptotic expansion of the first term is straightforward:
\begin{align}
  &\sum_{j=0}^{n}2^{j+1}(2j-1)!!\,T^j\int_0^{\be/4}\frac{y^j}{j!}\ln(1-e^{-y})\diff y\notag\\
  &=\sum_{j=0}^{n}2^{j+1}(2j-1)!!\,T^j\int_0^{\infty}\frac{y^j}{j!}\ln(1-e^{-y})\diff y
    +O(\be^ne^{-\be/4})\notag\\
  &=-\sum_{j=0}^{n}2^{j+1}(2j-1)!!\,\ze(j+2)T^j+O(\be^ne^{-\be/4})\,.
    \label{first}
\end{align}
We claim that the second term in the previous equation for $g(T)$ is $O(T^{n+1})$. This is easily
proved by integrating by parts $n+1$ times:
\begin{align*}
  &\be^{n+1}\int_0^{\be/4}\ln(1-e^{-y})\frac{\diff^{n+1}}{\diff y^{n+1}}h(Ty)\,\diff y\\
  &=\sum_{k=0}^n\left.(-1)^{n-k}\be^{n+1-k}\vp^{(n-k)}(y)h^{(k)}(Ty)\right|^{\be/4}_0\\
  &\hphantom{={}}+(-\be)^{n+1}\int_0^{\be/4}\vp^{(n+1)}(y)h(Ty)\,\diff y,
\end{align*}
with $\vp(y):=\ln(1-e^{-y})$. The boundary terms vanish at $y=0$,
since $h(z)=O(z^{2n+2})$. On the other hand,
\[
  \be^{n-k+1}\vp^{(n-k)}(\be/4)h^{(k)}(1/4)=O(\be^{n-k+1}e^{-\be/4})
\]
for $k=0,\dots,n$, as the first $n$ derivatives of $h$ are bounded. Thus the boundary terms are
$O(\be^{n+1}e^{-\be/4})$, while
\[
  \be^{n+1}\int_0^{\be/4}\vp^{(n+1)}(y)h(Ty)\,\diff y=O(T^{n+1}),
\]
since $h(Ty)=O((Ty)^{2n+2})$ and the integral $\int_0^{\infty}y^{2n+2}\vp^{(n+1)}(y)\diff y$ is
convergent. Putting all of the above together we obtain the asymptotic series
\[
  g(T)\sim-\sum_{k=0}^{\infty}2^{k+1}(2k-1)!!\,\ze(k+2)T^k,
\]
from which it follows that
\begin{equation}
  \label{fasHS}
  f_\pm(T)-f_\pm(0)\sim-\sum_{k=0}^{\infty}(m^{k+1}-1)2^{k+1}(2k-1)!!\,\ze(k+2)T^{k+2}\,.
\end{equation}
In particular, truncating the series after the first term and using Eq.~\eqref{disp} we obtain
Eq.~\eqref{fasympHS}.

Consider, finally, the HS-B chain with $\ga=0$, for which $\cE(x)=x(1-x/2)$. From Eqs.~\eqref{fpm}
and \eqref{fpmHS} it easily follows that
\[
  f_\pm(T)-f_\pm(0)=2\left(f_{\text{HS},\pm}(T/2)-f_{\text{HS},\pm}(0)\right)\,,
\]
whence we obtain the asymptotic series
\[
  f_\pm(T)-f_\pm(0)\sim-\sum_{k=0}^{\infty}(m^{k+1}-1)(2k-1)!!\,\ze(k+2)T^{k+2}\,.
\]
Again, the first term in this series is easily seen to yield Eq.~\eqref{fasymp}.

\section*{References}


%

\end{document}